\documentclass[12pt,preprint]{aastex}

\newcommand{\fer}{{\it Fermi}-LAT}
\newcommand{\wse}{{\it WISE}}
\newcommand{\swf}{{\it Swift}}
\newcommand{\igr}{{\it INTEGRAL}}

\newcommand{\gstrp}{{\it WGS}}
\usepackage{graphicx}
\usepackage{longtable}
\usepackage{rotating}
\usepackage{lscape}

\usepackage{lineno}
\usepackage{times}
    \usepackage{color}  

\newcommand{\ltsima} {$\; \buildrel < \over \sim \;$}
\newcommand{\gtsima} {$\; \buildrel > \over \sim \;$}
\newcommand{\lta} {\lower.5ex\hbox{\ltsima}}
\newcommand{\gta} {\lower.5ex\hbox{\gtsima}}

\slugcomment{version \today: fm}

\shorttitle{$\gamma$-ray blazar candidates among the unidentified \igr\ sources}
\shortauthors{F. Massaro, A. Paggi, R. D'Abrusco \& G. Tosti 2012}

\begin{document}

\linenumbers

\title{Searching for $\gamma$-ray blazar candidates among the unidentified \igr\ sources}
\author{F. Massaro\altaffilmark{1}, A. Paggi\altaffilmark{2}, R. D'Abrusco\altaffilmark{2} \& G. Tosti\altaffilmark{3,4}.}

\affil{SLAC National Laboratory and Kavli Institute for Particle Astrophysics and Cosmology, 2575 Sand Hill Road, Menlo Park, CA 94025}
\affil{Harvard - Smithsonian Astrophysical Observatory, 60 Garden Street, Cambridge, MA 02138}
\affil{Dipartimento di Fisica, Universit\`a degli Studi di Perugia, 06123 Perugia, Italy}
\affil{Istituto Nazionale di Fisica Nucleare, Sezione di Perugia, 06123 Perugia, Italy}

\begin{abstract}
The identification of low-energy counterparts for  $\gamma$-ray sources
is one of the biggest challenge in modern $\gamma$-ray astronomy. 
Recently, we developed and successfully applied a new association method  to recognize  $\gamma$-ray blazar candidates
that could be possible counterparts for the unidentified $\gamma$-ray sources above 100 MeV in the second {\it Fermi} { Large Area Telescope (LAT)} catalog (2FGL).
This method is based on the { Infrared (IR)} colors of the recent Wide-Field Infrared Survey Explorer (\wse) all-sky survey.
In this letter we applied our new association method to the case of unidentified \igr\ sources (UISs) listed in the 
fourth soft gamma-ray source catalog (4IC).
Only 86 UISs out of the 113 can be analyzed, due to the sky coverage of the \wse\ Preliminary data release.
Among these 86 UISs, we found that 18 appear to have a $\gamma$-ray blazar candidate within their positional error region.
Finally, we analyzed the \swf\ archival data available for 10 out these 18 $\gamma$-ray blazar candidates, 
and we found that 7 out of 10 are clearly detected in soft X-rays 
and/or in the optical-ultraviolet band. { We cannot confirm the associations between the UISs and the selected $\gamma$-ray
blazar candidates due to the discrepancies between the \igr\ and the soft X-ray spectra.
However, the discovery of the soft X-ray counterparts for the selected $\gamma$-ray blazar candidates 
adds an important clue to help understand their origin and to confirm their blazar nature.}
\end{abstract}


\section{Introduction}
\label{sec:intro}
One of the main scientific objectives of the \igr\ mission 
is performing a survey of the sky in a mostly unexplored region of the electromagnetic spectrum:
from the hard X-ray to the soft $\gamma$-ray band.
Since its launch in October 2002, \igr\ has used the unprecedented imaging capabilities of  IBIS 
\citep[Imager on Board INTEGRAL Spacecraft:][]{ubertini03} coupled with those of 
 ISGRI \citep[INTEGRAL Soft Gamma-Ray Imager;][]{lebrun03}.
Combining { data from} these two instruments, it is possible to generate images of the sky with a 
12 arcmin (Full Width Half Maximum, FWHM) resolution with typical source location accuracy of
$\sim$ 1-3 arcmin over a $\sim$19 degree (FWHM) field of view in the energy range 17--1000 keV.

The fourth soft $\gamma$-ray source catalog\footnote{http://irfu.cea.fr/Sap/IGR-Sources/} \citep{bird10}
(hereinafter 4IC) obtained with the IBIS $\gamma$-ray imager on
board the \igr\ satellite lists 723 hard X-ray/soft $\gamma$-ray sources.
In particular, the 4IC substantially increased the extragalactic sky coverage 
including both transients and faint persistent objects that can only 
be revealed with long exposure observations \citep{bird10}.

Several observations at low energies have been already performed to decrease the number of the unidentified \igr\ objects (UISs) 
\citep[see e.g.,][for optical and X-ray observations of UISs]{masetti08,masetti09,masetti10};
however, a considerable fraction of the 4IC sources are still completely unidentified.
According to the 4IC, there are 113 UISs,
corresponding to about 16\% of the whole catalog, and  178 other sources have uncertain classification.
The largest fraction (i.e., 35\%) of the associated \igr\ sources are Active Galactic Nuclei { (AGN)}, 
{ compared} to 31\%  identified as Galactic sources \citep{bird10}.
For comparison, the 58-month catalog of { observations with the} BAT hard X-ray detector 
\footnote{http://heasarc.nasa.gov/docs/swift/results/bs58mon/}
on board the \swf\ observatory, lists 1092 objects detected in the 14-195 keV energy range, 
with 86 unidentified hard X-ray sources listed \citep{cusumano10,baumgartner10}.

Recently, using the  \wse\ all-sky IR survey, we discovered that blazars, 
the largest known $\gamma$-ray class of AGN, 
can be separated from other extragalactic sources 
using  IR colors \citep[][hereinafter Paper I]{massaro11a}.
We used the magnitudes of the recent
\wse\ IR all-sky survey performed at 3.4, 4.6, 12, and 22 $\mu$m 
with an angular resolution of 6.1, 6.4, 6.5 \& 12.0 arcsec and
with 5$\sigma$ point source sensitivities achieving 0.08, 0.11, 1 and 6 mJy, 
in unconfused regions on the ecliptic, respectively. 
The absolute (radial) differences between \wse\ source-peaks and ``true" astrometric positions 
anywhere on the sky are no larger than $\sim$ 0.50, 0.26, 0.26, and 1.4 arcsec in the
four \wse\ bands, respectively \citep{cutri11}\footnote{http://wise2.ipac.caltech.edu/docs/release/prelim/expsup/sec2\_3g.html}.

Moreover, we investigated a sample of blazars detected by \wse\ and associated with \fer\ sources
to characterize their IR-$\gamma$-ray properties \citep[][hereinafter Paper II]{dabrusco12}.
This was the first step to develop a new association method for the unidentified $\gamma$-ray sources,
able to recognize if there is a $\gamma$-ray blazar candidate within their positional error region 
\citep[][hereinafter Paper III and Paper IV, respectively]{massaro12a,massaro12b}.
With this new IR diagnostic tool, we searched for $\gamma$-ray blazar candidates within the unidentified $\gamma$-ray source sample of the { 2FGL} $\gamma$-ray catalog, and for the first time 
we have been able to provide a candidate counterpart for 187 out of 313 unidentified $\gamma$-ray sources analyzed,
having the same IR properties as the $\gamma$-ray emitting blazars (see Paper IV).

In this letter, we apply this association procedure to { test whether} there is a possible $\gamma$-ray blazar 
candidate 
for the UISs using their \wse\ IR colors.
For the selected $\gamma$-ray blazar candidates, we also search \swf\ pointed observations
 for the presence of an optical-UV and/or X-ray counterpart.
This letter is organized as follows: in Section~\ref{sec:sample} we describe the UIS
sample selected for our investigation; in Section~\ref{sec:method} we illustrate the basic details of
our new association method, that, in Section~\ref{sec:uis}, we apply to the UISs. 
Section~\ref{sec:swift} is devoted to the optical-UV-X-ray counterparts in the \swf\ observations available.
Our results are discussed in Section~\ref{sec:results}.

\section{The sample selection}
\label{sec:sample}
In the 4IC there are 113 sources that are completely unidentified \citep[i.e., Type =?, according to Table 3 in][]{bird10},
while there are 97 sources that are indicated as unidentified transients 
\citep[i.e., Type =?,T, according to Table 3 in][]{bird10}.
In addition, within the 4IC, there are also 32 AGN sources of uncertain type
\citep[i.e., Type =AGN?][]{bird10} and another 49 objects with uncertain classification.

In this letter, we only considered the 86 UISs { out of 113} listed in the 4IC
that lie in the portion of the sky covered by the \wse\ Preliminary Source catalog.

\section{The \gstrp\ association method}
\label{sec:method}
In Paper III, using a subsample of the ROMA-BZCAT blazar catalog \citep{massaro09,massaro10,massaro11b}, 
detected by \wse\ and associated with \fer\  sources (Papers II and III), 
we presented the parametrization of the \wse\ $\gamma$-ray strip (\gstrp) based on the {\it strip parameter} $s$. 
This parameter, ranging between 0 - 1, provides an estimate of the
distance between the \gstrp\ and the location of a generic \wse\ source in the IR color parameter space,
and it is weighted for the errors on all the IR colors.
We distinguished between \wse\ sources that lie in the subregion of the \gstrp\
occupied by the BZBs and BZQs using the $s_b$ and $s_q$ parameters separately (Paper III).

In Paper IV, we presented the association method based on the \gstrp\ parametrization.
For each unidentified $\gamma$-ray source we defined the {\it searching region} corresponding 
to a circular region of radius $R$ equal to the semi-major axis of the elliptical source location region 
at 99.999\% confidence level, centered on the $\gamma$-ray position given in the 2FGL catalog\citep{Nolan12}.

We calculate the IR colors for every \wse\ source that lies within the {\it searching region}
as well as their $s_b$ and $s_q$ parameters.
Given the distributions of generic \wse\ sources in random regions of the sky,
we distinguish three classes of $\gamma$-ray blazar candidates on the basis of their $s_b$ and/or $s_q$ values:
\begin{itemize}
\item{class A: \wse\ sources with 0.24 $<s_b<$ 1.00 and 0.38 $<s_q<$ 1.00;}
\item{class B: \wse\ sources with 0.24 $<s_b<$ 1.00 or 0.38 $<s_q<$ 1.00;}
\item{class C: \wse\ sources with 0.10 $<s_b<$ 0.24 and 0.14 $<s_q<$ 0.38.}
\end{itemize}
All the \wse\ sources with $s_b<$0.10 or $s_q<$0.14 are considered {\it outliers} of the \gstrp.
Sources of class A are the rarest with respect to the other classes (Paper IV).

Our association procedure consists in
ranking all the \wse\ sources within the {\it searching region} of an unidentified $\gamma$-ray source
as described above and indicating as a $\gamma$-ray blazar candidate the positionally closest source belonging to the highest class.
Our association procedure provides a completeness of 87\% based on the {\it a posteriori} re-association of the 
ROMA-BZCAT blazars, detected by \wse\ and associated with \fer\ sources.

\section{$\gamma$-ray blazar candidates among the unidentified \igr\ sources}
\label{sec:uis}
We applied our new association method to the case of the 86 UISs selected above.
This { process} allows us to verify if there is a $\gamma$-ray blazar candidate within the positional error region of each UIS analyzed.

We considered a {\it searching region} with radius equal to the position error 
at 90\% confidence level, as reported in the 4IC catalog;
then, we estimated the IR \wse\ colors for all the sources that lie within the {\it searching region}. 

Running our association procedure, we found that 68 out of 86 UISs have only outliers of the \gstrp\ lying in their {\it searching regions},
while within the remaining 18 UISs we found 4 $\gamma$-ray blazar candidates of class A, 12 of class B and 2 of class C.
In Table 1, we present the list of $\gamma$-ray blazar candidates found for the 18 UISs together with their IR colors, as well as the $s_b$ and $s_q$ parameters.
We also estimated the probability to find a generic \wse\ source with the same $s$ values in 36 random circular regions of the \wse\ sky
having the same radius $R$ of the {\it searching regions}. We found that this is smaller than 10$^{-4}$.
We note that the positional accuracy of the UISs 
is { a least} order of magnitude better than that of the unidentified $\gamma$-ray sources in 2FGL.

Summarizing our results, we found 18 \wse\ $\gamma$-ray blazar candidates that could be candidate counterparts 
of the corresponding UISs responsible for the hard X-ray emission detected by \igr.

\section{$\gamma$-ray blazar candidates observed by \swf}
\label{sec:swift}
We found that among the $\gamma$-ray blazars selected according to our association procedure,
10 candidates out of 18 have at least one \swf\ pointed observation.
In addition, none of these $\gamma$-ray blazar candidates has a $\gamma$-ray counterpart in the 2FGL.
We reduced and analyzed these \swf\ observations to verify if these \wse\ $\gamma$-ray blazar candidates 
have an optical-UV or soft X-ray counterpart.
Here we report the data reduction and analysis procedures used in our \swf\ data analysis.
The comparison between the \swf\ and the \wse\ images will be presented in Section~\ref{sec:results}.

\subsection{UVOT data analysis}
\label{sec:uvot}
We followed the same { UV-Optical Telescope (UVOT)} reduction procedure described in \citep{tramacere07,massaro08a}
consequently, here we report only the basic details.

Several filter combinations are available for UVOT observations; however, 
we note that not all the optical and UV data are available for each source. 
The detection algorithm $UVOTDETECT$ was used 
to confirm the presence of the optical-UV counterpart of the $\gamma$-ray blazar candidates.
We then performed the photometric analysis using the $UVOTSOURCE$ tool.
Counts were extracted from a 6$''$ radius aperture in the $V$, $B$,
and $U$ filters and from a 12$''$ radius aperture for the other UV filters
($UVW1$, $UVM2$, and $UVW2$), to properly take into account the wider 
Point Spread Function in these bandpasses.
The count rate was corrected for coincidence loss, and the background
subtraction was performed by estimating its level in an offset circular region 
at 20$''$ from the source.

The correction for the interstellar reddening was obtained assuming the 
$E(B-V)$ values from Schlegel et al. (1998) and the corrections described in Cardelli et al. (1989), while
the fluxes were derived with the same conversion factors given by Giommi et al. (2006). 

\subsection{XRT data analysis}
\label{sec:xrt}
The X-Ray Telescope (XRT) data reduction used in the following is also the same one
described in \citep{tramacere07,massaro08b,massaro11c}; here we only report the basic details.

The XRT data analysis has been performed with the $XRTDAS$ software 
(v.~2.1), developed at the ASI Science Data Center (ASDC) and 
distributed within the HEAsoft package (v.~6.10.0). 

Event files were calibrated and cleaned 
with standard filtering criteria using the \textsc{xrtpipeline} task, 
combined with the latest calibration files available in the \swf\ $CALDB$ distributed by HEASARC. 
Only events in the energy range 0.3--10 keV with grades 0--12 were used.
When more than a single \swf\ pointing of each source has been performed and is available within the \swf\ archive, 
we { combined} several low S/N observations, because the 
the co-added X-ray image increases significantly the source detection.
No signatures of pile-up were found in our XRT observations.

Given the low exposure of the \swf\ observations it was not possible to carry out a detailed spectral analysis,
so unless stated otherwise we used the detection algorithm $detect$,
a tool of the $XIMAGE$ package for all the \swf\ observations.
The $detect$ algorithm locates the X-ray point sources using a sliding-cell method taking into account 
the average background intensity.
The position and intensity of each detected source is calculated using a box
whose size maximizes the signal-to-noise ratio. 
This detection algorithm has been extensively used in the 
Swift serendipitous survey in deep XRT gamma-ray burst fields \citep[see also][for additional details]{puccetti11}.
Statistical and systematic uncertainties on count rates are added quadratically.

Finally, we measured the net count rates for each detected soft X-ray source and we converted them
into fluxes assuming a power-law spectrum with spectral index 1 
and using WEBPIMMS\footnote{http://heasarc.nasa.gov/Tools/w3pimms.html}.

\section{Results on the \swf\ analysis}
\label{sec:results}
Applying our new association procedure developed for the unidentified $\gamma$-ray sources of \fer\ to the UISs,
we found that 18 sources out of the 86 analyzed have a $\gamma$-ray blazar candidate as possible counterparts.
We note that this new association method proposed for the unidentified $\gamma$-ray sources of \fer\ 
{ does}
not have the same efficiency when { applied to}
soft $\gamma$-rays and/or hard X-rays.
In fact, in the \fer\ energy range (i.e., 30 MeV - 10 GeV) blazars are the largest known $\gamma$-ray population \citep{Nolan12},
while the hard X-ray band is generally dominated by the emission of different classes of AGN, such as Seyfert galaxies,
which constitute $\sim$ 24\% in comparison with the 2.4\% for blazars 
already associated in the 4IC.
This implies that the \wse\ $\gamma$-ray blazar candidates are not necessarily the low-energy counterparts of the UISs.

{ The relation between the IR spectral shape in the \wse\ energy range and that in the $\gamma$-rays
is { based on}
our association method (e.g., Paper II).
On the other hand, in  hard X-rays there is not yet evidence of a  
correlation between the IR and the X-ray emission of blazars, thus 
the eventual association between the \wse\ $\gamma$-ray blazar candidate and the 
UISs is less robust that in the case of the \fer\ sources.}

\begin{figure}[]
\includegraphics[height=6.4cm,width=8.8cm,angle=0]{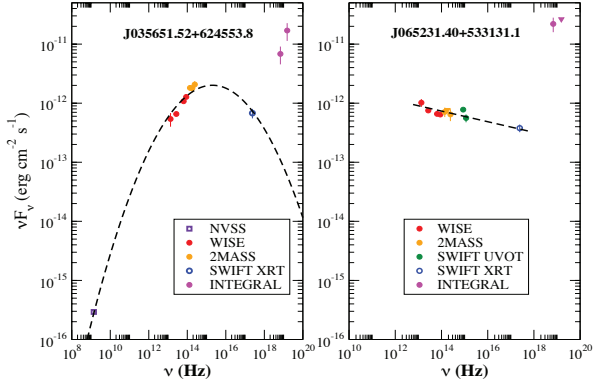}
\caption{The SEDs of two example of \wse\ $\gamma$-ray blazar candidates: J035651.52+624553.8 and J065231.40+533131.1.
The dashed line in the { left}
panel is the typical log-parabolic model adopted to describe the non-thermal SED of J035651.52+624553.8
while in the case of J065231.40+533131.1 a simple power-law, over 6 orders of magnitude, has been used.
As described in Section~\ref{sec:results} there is discrepancy between the XRT fluxes and those of \igr\ that do not support the 
blazar association of the UIS.}
\label{fig:seds}
\end{figure}

For 10 out of the18 \wse\ $\gamma$-ray blazar candidates, we also found optical-UV and X-ray observations available in the \swf\ archive
that could be helpful to verify if they are the low-energy counterparts of the UISs.
We found that 7 out of the 10 \wse\ $\gamma$-ray blazar candidates in the \swf\ archive have a clear counterpart
in  X-rays and in the optical-UV bands, showing a typical { Spectral Energy Distribution (SED)} dominated by non-thermal emission, as { for} the two cases shown in Figure~\ref{fig:seds},
where J035651.52+624553.8 has also a radio counterpart at 1.12 arcsec from the \wse\ position.
In Table 1, we report the \igr\ name together with the \wse\ $\gamma$-ray blazar candidates, the J2000 coordinates RA and DEC,
the distance between the \wse\ source and the \igr\ position in arcsec, the \wse\ colors (i.e., $c_{12}=[3.4]-[4.6]$, $c_{23}=[4.6]-[12]$, $c_{23}=[12]-[22]$),
the $s_b$ and the $s_q$ derived from our \gstrp\ method, the \swf\ UVOT detections and the \swf\ XRT detections 
with the X-ray counts in the soft (0.3-1 keV) and
in the hard (i.e., 1-10 keV) bands together with the X-ray hardness ratio $HR$ derived from the net number of counts.

For remaining three \wse\ $\gamma$-ray blazar candidates we did not find
a clear counterpart in \swf\ observations. This result could be due to the short exposures of the archival observations. 

\begin{figure}[]
\includegraphics[height=6.6cm,width=8.0cm,angle=0]{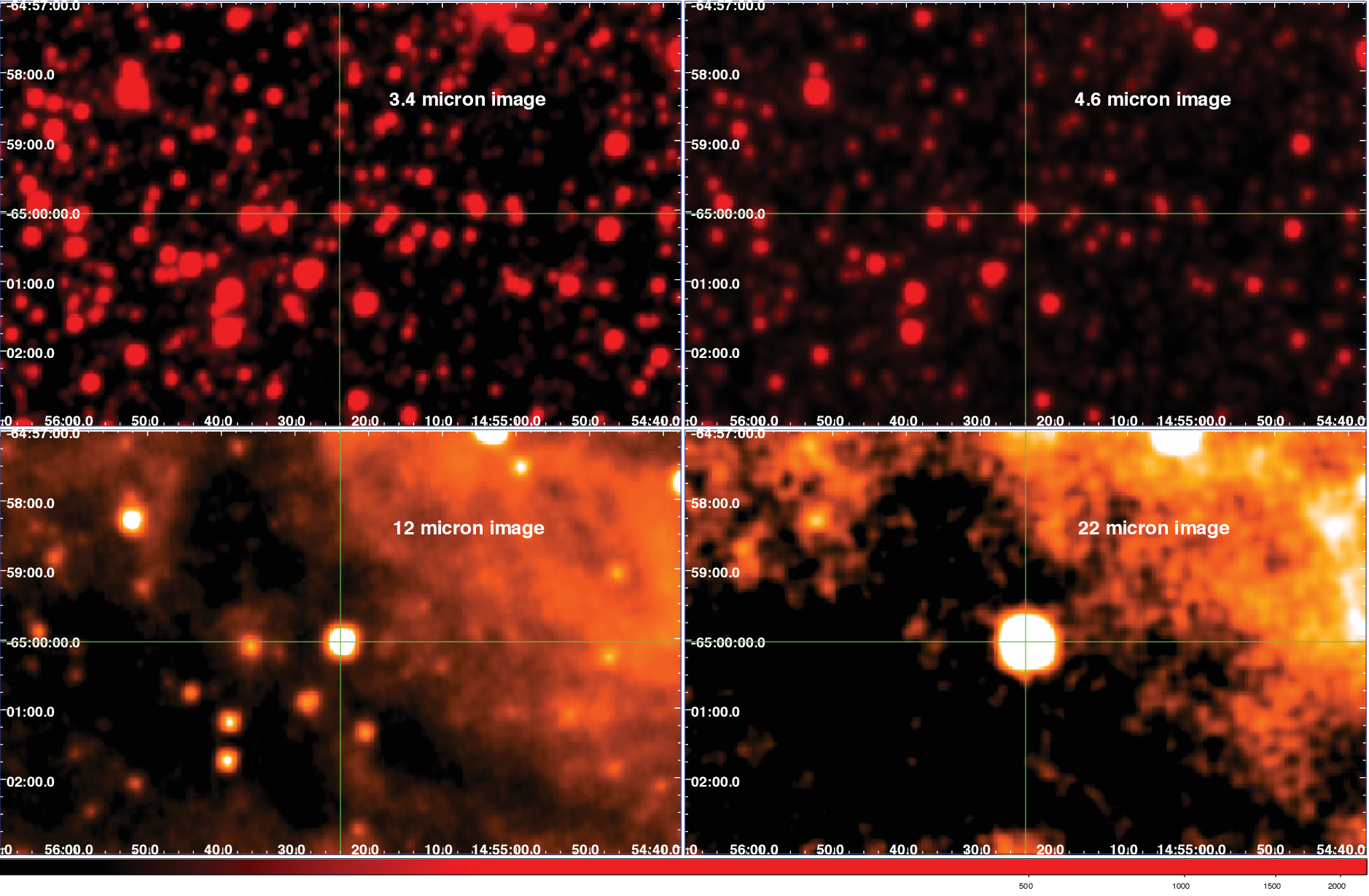}
\caption{The  \wse\ IR images at 3.4$\mu$m, 4.6$\mu$m, 12$\mu$m, 22$\mu$m, respectively for the { Field of View (FOV) for}  IGRJ14549$-$6459,
centered on the position of the $\gamma$-ray blazar candidate.
The \wse\ $\gamma$-ray blazar candidate is highlighted with the green cross in the center of the \wse\ images. It is clear that the source selected with our association method
is the only one detected in all  4 \wse\ bands.}
\label{fig:igr14549_wise}
\end{figure}

However, we note that in the above 10 candidates, the \swf\ XRT flux is not in agreement with the extrapolation of the \igr\ spectrum,
which is generally one order of magnitude larger than the \swf\ XRT estimate.
{ This discrepancy is not sufficient to exclude the blazar association of the UISs because blazars exhibit rapid X-ray variability;
however, the it could suggest that
the blazar is not the most probable low-energy counterpart for the UISs,}
in agreement with the fact that they are not the dominant class of AGN in the hard X-rays.
We note that the $\gamma$-ray blazar candidates found with our method are \wse\ sources, detected in all four \wse\ bands, in particular at 22 $\mu$m 
as the case of IGRJ14549$-$6459 shown in Figure~\ref{fig:igr14549_wise}, for which the \wse\ candidate counterpart appear to have the IR colors of blazars. 

\begin{figure}[]
\includegraphics[height=6.6cm,width=18.0cm,angle=0]{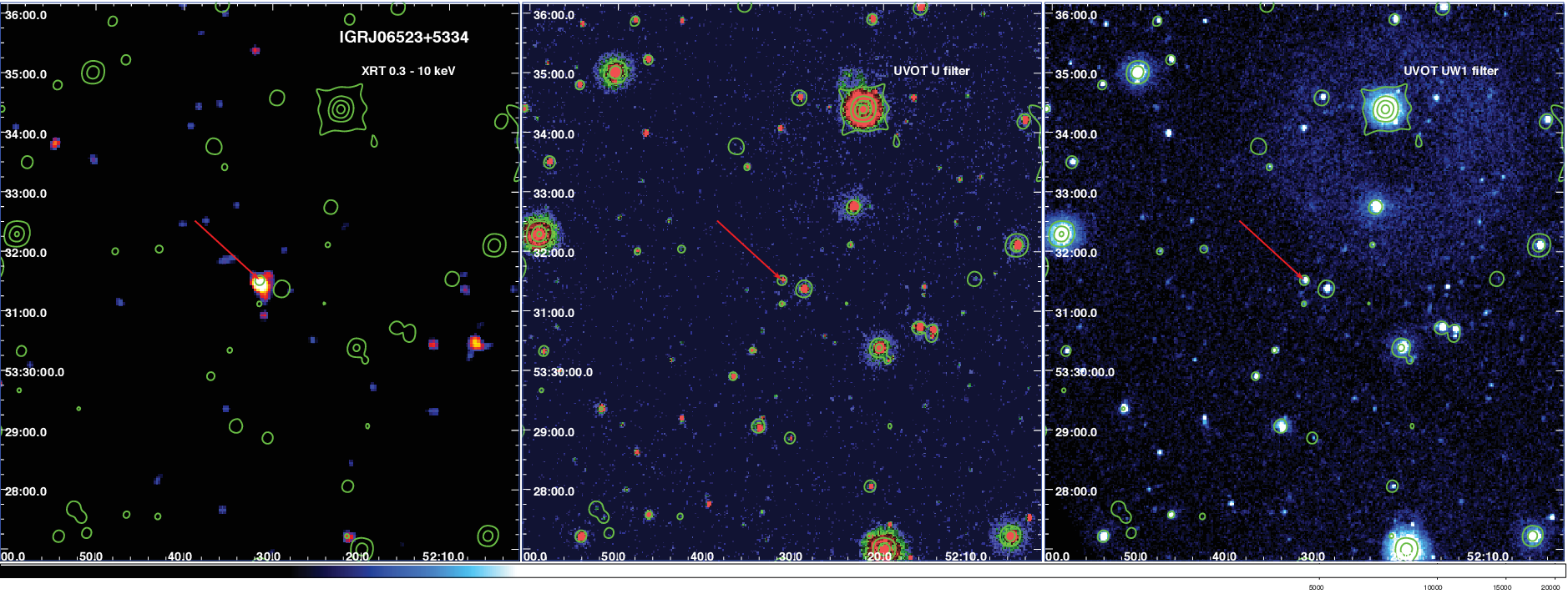}
\caption{The \wse\ 3.4 $\mu$m IR contours { (green)} overlaid on the \swf\ optical-UV and X-ray images, for the FOV of IGRJ06523+5334,
centered on the position of the selected \wse\ blazar candidate.
It is clear that the \wse\ blazar candidate (red arrow) has a clear counterpart in the soft X-rays and in the optical-UV bands.}
\label{fig:igr06523}
\end{figure}

\begin{figure}[]
\includegraphics[height=6.6cm,width=18.0cm,angle=0]{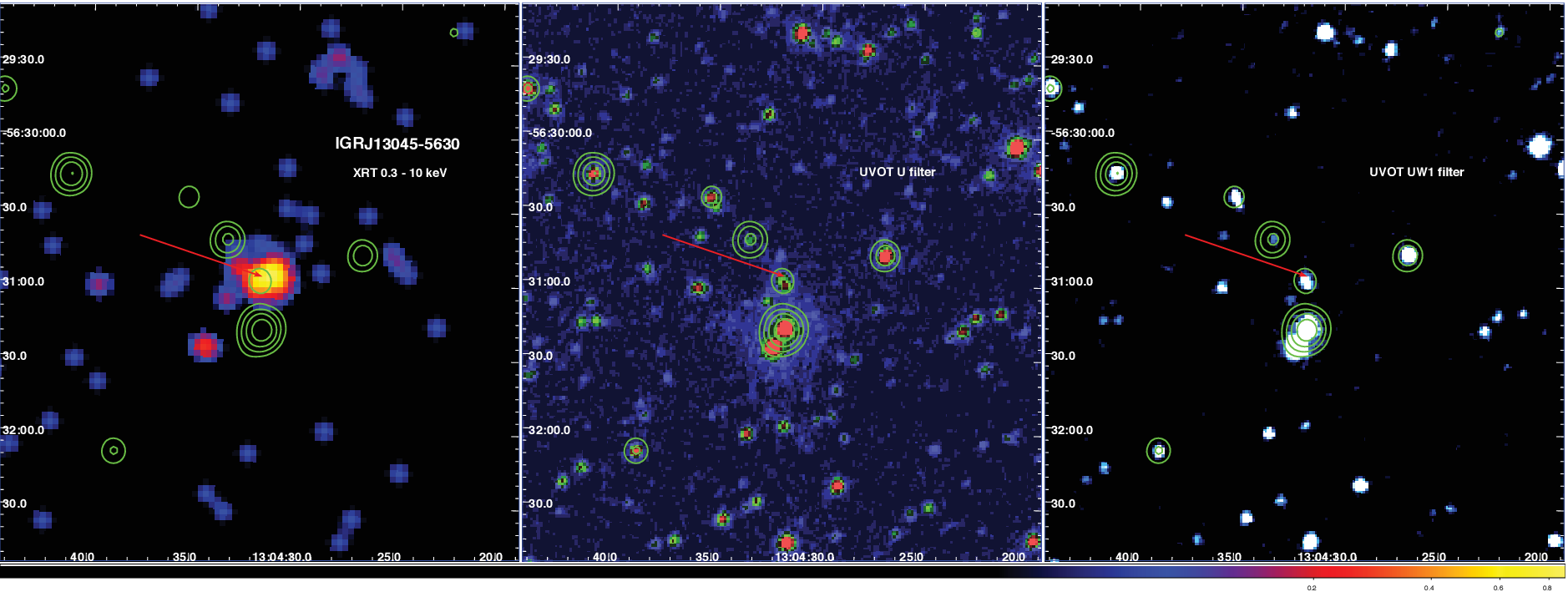}
\caption{Same of Figure~\ref{fig:igr06523} for the IGRJ 13045$-$5630 FOV.}
\label{fig:igr13045}
\end{figure}

In Figure~\ref{fig:igr06523} and Figure~\ref{fig:igr13045}, we also show the comparison between the soft X-ray and the optical-UV images 
of \swf\ with the \wse\ IR data, for the FOV of two examples of UISs centered on the positions of our $\gamma$-ray blazar candidates: 
IGRJ06523+5334 and IGRJ13045$-$5630, respectively.

\section{Summary}
We applied our new association method successfully used for the unidentified $\gamma$-ray sources in the 2FGL to the UISs
to test if it is possible to find \wse\ blazar counterparts at low energies responsible for the hard X-ray emission detected by \igr within the {\it searching regions} of the UISs.

We found that 18 out of 86 UISs analyzed clearly have a blazar counterpart within the {\it searching regions},
and for 10 of them also \swf\ archival observations are available.
However, for the latter 10 sources in \swf\ we did not find a good agreement between the \swf\ X-ray flux and the one estimated by the extrapolation of the \igr\ spectrum.
Thus, we are not able to confirm if the \wse\ $\gamma$-ray blazar candidates found with our method could be associated with the UISs considered.

On the other hand, one crucial result arises from our analysis. We found that the \wse\ $\gamma$-ray blazar candidates selected from our method 
in these serendipitous \swf\ observations of the UIS fields of view have clear optical and/or UV and soft X-ray counterparts.
This is strongly in agreement with the expectations driven by their blazar nature.
It is worth noting that J035651.52+624553.8 has also a radio counterpart and
in addition, the SEDs of these \wse\ $\gamma$-ray blazar candidates are in agreement with a non-thermal shape over several orders of magnitude.

Finally, we remark that future follow up observations, { in particular spectroscopic optical data,}
are necessary to clarify the nature of the \wse\ $\gamma$-ray blazar candidates 
and consequently the nature of the UISs \citep[see e.g.,][]{masetti08,masetti09,masetti10}.

\acknowledgements
We thank the anonymous referee for the his/her comments.
We are grateful to D. J. Thompson for all his comments helpful toward improving our presentation. 
F. Massaro is grateful to H. Smith, J. Grindlay, M. Ajello, E. Bottaccini for their helpful discussions.
The work at SAO is supported in part by the NASA grant NNX10AD50G and NNX10AD68G.
R. D'Abrusco gratefully acknowledges the financial 
support of the US Virtual Astronomical Observatory, which is sponsored by the
National Science Foundation and the National Aeronautics and Space Administration.
TOPCAT\footnote{\underline{http://www.star.bris.ac.uk/$\sim$mbt/topcat/}} 
\citep{taylor2005} and SAOImage DS9 were used extensively in this work. 
Part of this work is based on archival data, software or on-line services provided by the ASI Science Data Center.
This publication makes use of data products from the Wide-field Infrared Survey Explorer, 
which is a joint project of the University of California, Los Angeles, and the Jet Propulsion Laboratory/California Institute of Technology, 
funded by the National Aeronautics and Space Administration.

{}

\begin{landscape}
\begin{table}
\tiny
\begin{tabular}{|llccccccccccc|}
\hline
\igr                  & \wse         & RA          & DEC       & distance   & $c_{1}$   & $c_{2}$   & $c_{3}$   & $s_b$ & $s_q$ & UVOT/ XRT  & counts & counts \\
name             & name        &  (deg) &  (deg)  & arcsec      &                   &                  &                  &            &            & detec.  & 0.3-1 keV & 1-10 keV \\
\hline
\noalign{\smallskip}			         
class A sources & & & &  & & & & & & & & \\
\noalign{\smallskip}			         
\hline
\noalign{\smallskip}			         
IGRJ04442+0450 & J044415.86+045126.6 &  71.07 &   4.86 &  88.17 & 1.17(0.03) & 3.06(0.03) & 2.81(0.05) & 0.43  & 0.75  &   -   & - & - \\
IGRJ06523+5334 & J065231.40+533131.3 & 103.13 &  53.53 & 219.42 & 1.02(0.04) & 3.01(0.06) & 2.45(0.13) & 0.30  & 0.43  &   n/y   & 18 & 41 \\
IGRJ14549-6459 & J145523.80-650002.5 & 223.85 & -65.00 & 212.20 & 1.09(0.03) & 2.71(0.03) & 2.43(0.03) & 0.91  & 0.92  &   n/y  & - & 38  \\
IGRJ16413-4046 & J164122.31-404714.5 & 250.34 & -40.79 &  28.13 & 0.73(0.04) & 2.05(0.03) & 1.36(0.03) & 0.94  & 0.39  &   n/n   & - & - \\
\hline
\noalign{\smallskip}			         
class B sources & & & &  & & & & & & & & \\
\noalign{\smallskip}			         
\hline
\noalign{\smallskip}			         
IGRJ03502-2605 & J035018.94-260423.6 &  57.58 & -26.07 & 116.77 & 1.12(0.04) & 2.51(0.07) & 2.33(0.22) & 0.26  & 0.31  &   n/ y   & 10 & 2 \\
IGRJ03564+6242 & J035651.52+624553.8 &  59.21 &  62.76 & 264.69 & 0.87(0.04) & 2.40(0.06) & 1.93(0.22) & 0.35  & 0.17  &   n/y   & 19 & 12 \\
IGRJ07225-3810 & J072228.14-381457.6 & 110.62 & -38.25 & 293.05 & 1.06(0.05) & 2.68(0.09) & 1.89(0.20) & 0.26  & 0.24  &   -   & - & - \\
IGRJ13045-5630 & J130431.77-563058.5 & 196.13 & -56.52 &  58.56 & 0.94(0.03) & 3.30(0.03) & 2.71(0.03) & 0.00  & 0.69  &   n/y   & 4 & 40 \\ 
IGRJ13107-5626 & J131037.06-562654.3 & 197.65 & -56.45 &  29.06 & 1.29(0.03) & 2.78(0.03) & 2.26(0.04) & 0.00  & 0.84  &   -   & - & - \\
IGRJ15293-5609 & J152900.40-560830.4 & 232.25 & -56.14 & 149.87 & 0.82(0.06) & 2.44(0.06) & 2.39(0.07) & 0.43  & 0.29  &   -   & - & - \\
IGRJ15311-3737 & J153051.78-373457.1 & 232.72 & -37.58 & 211.38 & 0.87(0.03) & 2.17(0.03) & 2.05(0.06) & 0.70  & 0.29  &   y/y  & 45 & 220 \\
IGRJ16560-4958 & J165551.96-495732.3 & 253.97 & -49.96 &  59.51 & 0.75(0.05) & 2.28(0.04) & 2.05(0.08) & 0.52  & 0.22  &   -   & - & - \\
IGRJ17314-2854 & J173111.38-285701.8 & 262.80 & -28.95 & 180.76 & 0.34(0.03) & 1.25(0.02) & 0.88(0.03) & 0.00  & 0.44  &   -   & - & - \\
IGRJ17448-3232 & J174440.89-323155.8 & 266.17 & -32.53 &  89.35 & 0.64(0.04) & 1.88(0.03) & 1.19(0.05) & 0.45  & 0.30  &   n/n  & - & - \\
IGRJ19552+0044 & J195504.07+004421.0 & 298.77 &   0.74 & 106.47 & 1.04(0.05) & 2.72(0.08) & 2.18(0.25) & 0.29  & 0.29  &   -   & - & - \\
IGRJ20450+7530 & J204522.41+753057.4 & 311.34 &  75.52 &  90.73 & 0.87(0.04) & 2.29(0.08) & 2.30(0.29) & 0.27  & 0.16  &   y/y   & 17 & 52 \\
\hline
\noalign{\smallskip}			         
class C sources & & & &  & & & & & & & & \\
\noalign{\smallskip}			         
\hline
\noalign{\smallskip}			         
IGRJ13550-7218 & J135453.52-721422.4 & 208.72 & -72.24 & 217.52 & 1.12(0.07) & 2.49(0.10) & 2.37(0.28) & 0.17  & 0.22  &   n/n   & - & - \\
IGRJ16388+3557 & J163901.61+355510.7 & 249.76 &  35.92 & 200.60 & 1.07(0.05) & 2.75(0.10) & 2.64(0.26) & 0.19  & 0.23  &   -   & - & - \\
\hline
\noalign{\smallskip}			           
\end{tabular}\\
~\\
Col. (1) \igr\ name\\
Col. (2) \wse\ blazar candidates\\
Cols. (3, 4) the J2000 coordinates RA and DEC\\
Col. (5) the distance between the \wse\ source and the \igr\ position in arcseconds\\
Cols. (6,7,8) the \wse\ colors (i.e., $c_{1}=[3.4]-[4.6]$, $c_{2}=[4.6]-[12]$, $c_{2}=[12]-[22]$); the 1$\sigma$ errors are reported in parenthesis.\\
Cols. (9,10) the $s_b$ and the $s_q$ derived form our WGS method\\
Cols. (11,12) the \swf\ UVOT detections and the \swf\ XRT detections\\
Cols. (13,14) the X-ray counts in the soft (0.3-1 keV) and in the hard (i.e., 1-10 keV) band, respectively\\
\end{table}
\end{landscape}

\end{document}